# Simulation of phonon-assisted band-to-band tunneling in carbon nanotube field-effect transistors

Siyu Koswatta and Mark S. Lundstrom

School of Electrical and Computer Engineering, Purdue University, West Lafayette, IN

47907-1285

M. P. Anantram

NASA Ames Research Center, T27-A1, Moffett Field, CA 94035-1000

Dmitri E. Nikonov[a]

Technology and Manufacturing Group, Intel Corp., SC1-05, Santa Clara, CA 95052

Electronic transport in a carbon nanotube (CNT) metal-oxide-semiconductor field effect transistor (MOSFET) is simulated using the non-equilibrium Green's functions method with the account of electron-phonon scattering. For MOSFETs, ambipolar conduction is explained via phonon-assisted band-to-band (Landau-Zener) tunneling. In comparison to the ballistic case, we show that the phonon scattering shifts the onset of ambipolar conduction to more positive gate voltage (thereby increasing the off current). It is found that the subthreshold swing in ambipolar conduction can be made as steep as 40mV/decade despite the effect of phonon scattering.

05.60.+w, 72.10.-d, 72.80.Rj, 73.50.-h, 73.40.Gk, 85.30.Tv

---

[a] Electronic address: dmitri.e.nikonov@intel.com.

Electronic transport in conventional metal-oxide-semiconductor field-effect transistors (MOSFETs) is limited at room temperature primarily by surface roughness and phonon scattering. However carbon nanotubes, due to their high mobility, are believed to be operating in the ballistic regime under low source-drain bias. Carbon nanotube field effect transistors (CNTFET) [1] [2] [3] have been recently demonstrated. Non-equilibrium Green's function simulations of carbon nanotubes [4] [5] have been performed in the ballistic limit. It was demonstrated that the simulation results are in good agreement with experiments [6] at least over a wide range of parameters. Monte Carlo simulations [7] showed that in some regimes scattering produces only a small effect on current.

In this letter we introduce into the formalism a model of scattering due to electron-phonon interaction and explore the range of validity of the previous quasi-ballistic results. We show that the most dramatic difference occurs for the off-current, i.e. at the cross-over from the direct (over the barrier) conduction to ambipolar (tunneling) conduction [b]. The effects of band-to-band (Landau-Zener) tunneling [8] in CNTFETs have been experimentally investigated by Appenzeller et al. [9] Here we examine ambipolar conduction in the presence of phonon scattering. We explore the physics of the very steep subthreshold swing in ambipolar conduction as recently reported by Appenzeller et al. [9]. We describe regimes where it is preserved or destroyed by phonon scattering.

We apply the non-equilibrium Green's functions method, as described by Datta [10] to simulation of the carbon nanotubes transistors and extend the method of [4,5]. It involves solution of the equation for the Green's function, $G$, in the mode-space approach [11]

---

[b] Ambipolar conduction is a well-understood effect in CNT Schottky-barrier FETs. We show that it also occurs in CNT MOSFETs.

$$(EI - H - \Sigma)G(E) = I, \qquad (1)$$

where $E$ is the energy of quantum states, the imaginary part of self energy, $\Sigma = -i(\Sigma^{in} + \Sigma^{out})/2$, contains the contribution of respectively in- and –out scattering and contacts [c]. The real part of the self energy is obtained via the Hilbert transform [12] and corresponds to a shift of the energies of quantum states; we disregard it in this letter. The Hamiltonian, $H$, of the device includes the tight-binding coupling between carbon atoms with the bonding energy, $t = 3\text{eV}$, at the nearest neighbor distance, $a_{cc} = 0.142\text{nm}$, as well as the electrostatic potential energy. The latter is expressed via the potential, $V$, from a solution of the Poisson's equation with electric charges given by the electron and hole densities, $n$ and $p$, respectively; see [4,5] for the geometry of the device and the details of the computation. The carrier densities are obtained via integrals over energy of the electron and hole correlation functions,

$$G^n = G\Sigma^{in}G^+, \qquad G^p = G\Sigma^{out}G^+; \qquad (2)$$

they express the density of filled and unfilled states, respectively, over the energy and the coordinate. All the $G$-functions are matrices in the basis of the discrete points, $z$, along the nanotubes. The current at any point along the tube is[13]

$$I(z) = 4\frac{ie}{\hbar}\int \frac{dE}{2\pi}\left(G^{-1}(z, z+1, E)G^n(z+1, z, E) - G^{-1}(z+1, z, E)G^p(z, z+1, E)\right). \qquad (3)$$

---

[c] We disregard the effects of Coulomb blockade which are only significant for shorter channel length, thicker gate oxide, and lower gate dielectric constant, than in practically important devices considered in this paper.

We will be plotting the energy spectrum of the current, which is the expression under the integral.

In addition to commonly included contributions to self-energy from the source and drain electrodes, here we also introduce the contributions due to electron-phonon scattering

$$\Sigma^{in}(E) = R_{ph}(E,\omega)\left[(N_\omega+1)G^n(E+\hbar\omega) + N_\omega G^n(E-\hbar\omega)\right], \quad (4)$$

$$\Sigma^{out}(E) = R_{ph}(E,\omega)\left[(N_\omega+1)G^p(E-\hbar\omega) + N_\omega G^p(E+\hbar\omega)\right], \quad (5)$$

where the diagonal elements of the electron and hole correlation function are implied, the energy of a phonon mode is $\hbar\omega$, number of phonons in a mode, $N_\omega$, is given by the Bose-Einstein distribution. The electron-phonon coupling, $R_{ph}$, is proportional to the square of the deformation potential. In this letter for simplicity we take the fixed energy of phonons, 160meV, and set an independent of energy coupling constant which is related to the mean free path as

$$R_{ph} = \frac{t^2 3 a_{cc}}{2\lambda_{ph}}. \quad (6)$$

We model a CNTFET with a (n,m)=(13,0) carbon nanotube surrounded by a gate in a cylindrically symmetric geometry. The bandgap energy for this nanotube is, $E_g \approx 0.82\text{eV}$. The gate dielectric is HfO$_2$ with dielectric constant $\varepsilon = 16$ and thickness of 2nm. Source and drain are n-doped to linear density, $N_d$, the channel of 20nm in length is undoped. The doping density for the results in Figs. 1-3 is $N_d = 1.5 \cdot 10^9 / \text{m}$ and in Fig. 4 it is $N_d = 6 \cdot 10^8 / \text{m}$. This is to be compared with the linear density of carbon atoms of $N_{at} = 4n/(3a_{cc}) = 1.2 \cdot 10^{11} / \text{m}$ for this size of the nanotube. The contacts to the gate and the ends of source and drain are assumed to be infinite reflectionless reservoirs; bias voltages $V_g$ and $V_d$ are applied to them. We calculate the distribution of free carriers and the current along the carbon nanotubes. All the results in this letter are for the temperature of 300K. The simulation results for the electron occupation, $G^n(z,z,E)$, in states with a certain energy along the nanotube are presented in Fig. 1 for the ballistic case and in Fig. 2 for the case with phonon scattering. The source (left) and the drain (right, in Fig. 1) are filled with electrons up to the Fermi level. For the drain the Fermi level is shifted downwards due to the applied bias. Electrons occupy the states in the conduction band, but some electron density in the bandgap is also observed and corresponds to the evanescent tails of the wavefunctions. The negative gate bias forces the conduction and valence band in the channel to be at higher energy. Due to the relatively short length of the channel, discrete states corresponding to longitudinal confinement are formed in the valence band; they are clearly seen in the electron distribution. The current distribution over energies (not shown here) is position-independent in the ballistic case. Current flows by electron tunneling from the conduction band to the discrete states in the valence band and then by tunneling to the empty states in the conduction band of the drain (as evidenced by peaks of electron density there).

The changes brought by the inclusion of phonon scattering (Fig. 2) are as follows. Electron density is noticeable in the bandgap at approximately a value of energy of one phonon below the band edge in the source and drain. These are quasi-continuous virtual states caused by scattering, where electrons remain over the time sufficient to tunnel to the valence band. A discrete state (which would be better noticeable in the hole distribution $G^p(z,z,E)$) is formed at a value of energy of one phonon above the valence band in the channel. These phenomena bear a close similarity with those in resonant tunneling devices [14,15].

This insight into the processes governing conduction is confirmed by the current distribution (Fig. 3). It has a series of sharp peaks each corresponding to a discrete state in the channel. As the coordinate varies from the source to the drain, the current distribution shifts to lower energies due to emission of phonons. Uninterrupted lines of current distribution correspond to a fraction of electrons tunneling without scattering. Lines ending in the channel are explained by tunneling with a subsequent emission of a phonon. Lines starting in the bandgap show electrons first emitting phonons and then tunneling to lower energy discrete states. Apparently any sequence of two tunneling events and a number of phonon emissions contributes to current.

The effect of the phonon scattering is most visible in the dependence of the current on gate voltage, Fig. 4. The I-V characteristics in the ballistic approximation (dashed line) and with phonons (solid line) are close for large positive or negative values of gate voltage, in agreement with the results of a semiclassical Monte Carlo simulation [7]. For the positive gate voltage the current saturates when the potential barrier formed by the channel is low enough for all available carrier to flow without reflection. The current decreases with voltage with a subthreshold swing approximately 60mV/decade as expected for an over-the-barrier flow due to the tail of the Fermi distribution.

Current increases again at negative gate biases when tunneling conduction comes into play. It happens when the top of the valence band in the channel aligns in energy with the bottom of the conduction band in the source. Another condition is that unoccupied states in the drain are available at this energy. The current saturates at large negative voltages when sufficient number of discrete states are aligned to energies to transport carriers from source states are filled to the drain states are empty. Note that a large current ~ 1µA can be achieved through tunneling for sufficiently large drain bias and doping.

The difference between the ballistic limit and phonon scattering is most pronounced at small gate voltages where the cross-over between the over-the-barrier current and tunneling current. The onset of tunneling conduction is shifted to more positive gate biases. The reason for that is that conduction is possible with emission of a phonon when the top of the valence band in the channel is at a lower energy than the bottom of the conduction band in the source. This is in agreement with the fact that the shift in voltage is approximately equal to the energy of a phonon.

Our simulations show existence of a steep (<60mV/decade at room temperature) subthreshold swing in the IV curve in the ambipolar conduction region. The turn-on of this conduction with the change of the gate voltage happens due to overlap of the discrete states and a band edge. Thus it is not limited by the slope of the tail of the Fermi distribution. In Fig. 4, the swing in the ballistic regime is 25mV/decade around Vg=-0.3V. Similarly with phonon scattering the swing is 40mV/decade for Vg between -0.3 and -0.14V.

Note that this situation changes with the doping density, and consequently the Fermi level energy in the source relative to the band edge, $F_s$. It determines the on-off current ratio approximately as $\exp\bigl((E_g - F_s)/(k_B T)\bigr)$. For higher doping the on-off current ratio is lower, and the steep subthreshold swing exists in the ballistic approximation but disappears in the presence of phonon scattering. For example, for $N_d = 1.5 \cdot 10^9 / m$ a steep subthreshold swing is 26mV/decade at Vg=-0.1 is seen in the ballistic limit, while with phonon scattering it never exceeds 140mV/decade. The reason is that in all cases the turn-on of ambipolar conduction is less sharp than one expects from the ballistic calculations: smaller contribution to conduction exist at higher gate voltages when multi-phonon scattering becomes resonant with discrete states.

These results have significant consequences for the device design. One has to include phonon scattering to get the correct value of the gate voltage at which the current is minimal and to predict its exact value. In the presence of phonon scattering, one can obtain a sharp turn-on of current in ambipolar conduction, but at lower doping density and consequently smaller "on"-current.

In summary, we demonstrated that the ambipolar conduction in carbon nanotubes is ruled by phonon-assisted band-to-band tunneling. Phonon scattering shifts of the onset of ambipolar conduction to more positive gate voltage and thereby sets a lower limit for the off current. The steep subthreshold swing expected in tunneling conduction occurs for lower doping of the source and drain and is destroyed by phonon scattering for higher doping levels.

We acknowledge the support of this work by the National Science Foundation's Network for Computational Nanotechnology, NASA, and Intel Corporation.

**List of Figures**

**FIG. 1. (Color online) Distribution of electrons over energies in the carbon nanotube without phonon scattering; Vg=-0.55V, Vd=0.6V, source/drain doping $1.5*10^9$/m. White curves are the positions of the conduction (higher) and the valence (lower) band edges.**

**FIG. 2. (Color online) Distribution of electrons with phonon scattering, Rph=0.03(eV)$^2$; other parameters same as in Fig. 1.**

**FIG. 3. (Color online) Current spectrum (logarithmic scale) with phonon scattering. Parameters same as in Fig. 2.**

**FIG. 4. Current vs. gate voltage, for Vd=0.1: without scattering (dashed line); with phonon scattering (solid line with circles). The doping density is $6*10^8$/m.**

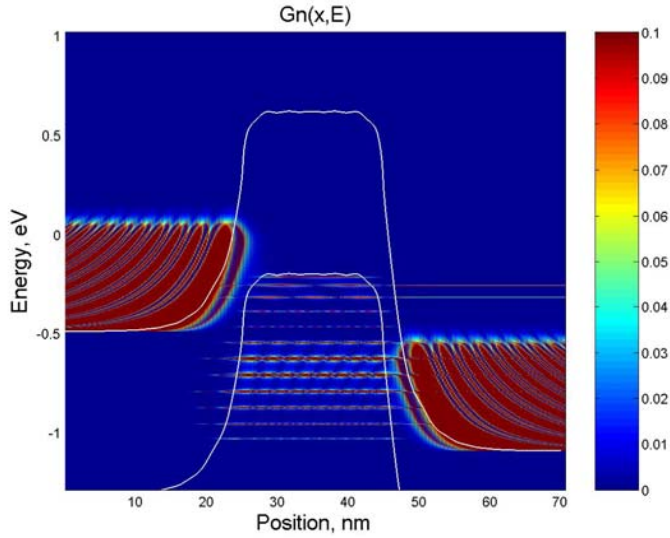

**FIG. 1. (Color online) Distribution of electrons over energies in the carbon nanotube without phonon scattering; Vg=-0.55V, Vd=0.6V, source/drain doping 1.5*10$^9$/m. White curves are the positions of the conduction (higher) and the valence (lower) band edges.**

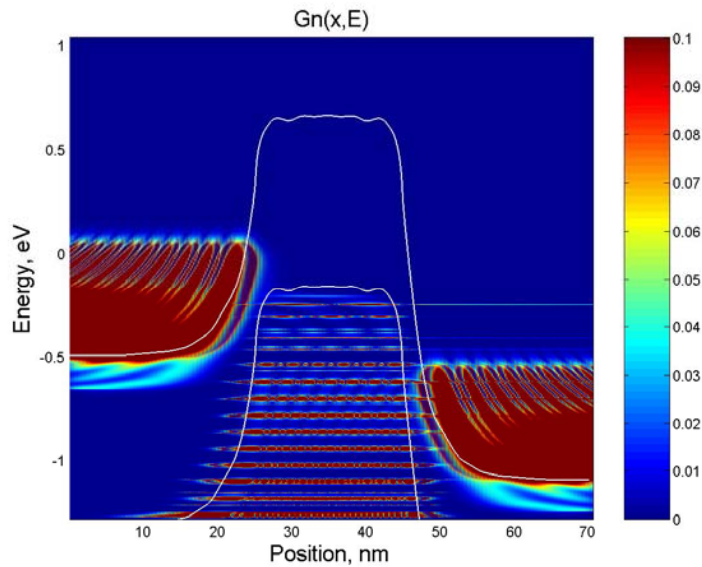

**FIG. 2. (Color online) Distribution of electrons with phonon scattering, Rph=0.03(eV)$^2$; other parameters same as in Fig. 1.**

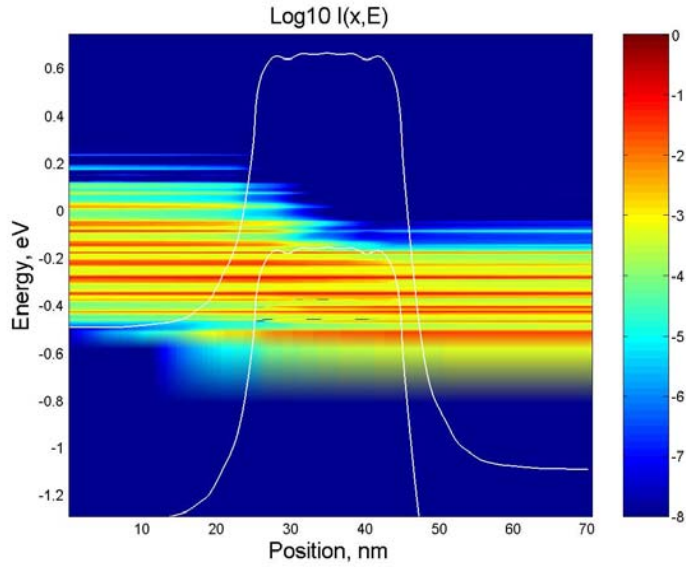

**FIG. 3. (Color online) Current spectrum (logarithmic scale) with phonon scattering. Parameters same as in Fig. 2.**

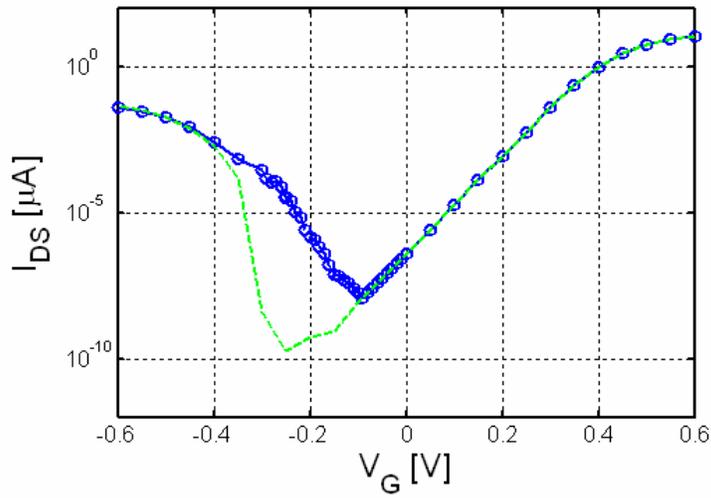

**FIG. 4. Current vs. gate voltage, for Vd=0.1: without scattering (dashed line); with phonon scattering (solid line with circles). The doping density is $6*10^8$/m.**

[1] S. Heinze, J Tersoff, R. Martel, V. Derycke, J Appenzeller, and Ph. Avouris, Phys. Rev. Lett. **89**, 106801 (2002).

[2] J. Appenzeller, J. Knoch, V. Derycke, R. Martel, S. Wind, and Ph. Avouris, Phys. Rev. Lett. **89**, 126801 (2002).

[3] A. Javey, J. Guo, Q. Wang, M. Lundstrom, and H. Dai, Nature **424**, 654 (2003).

[4] J. Guo, S. Datta, and M. Lundstrom, IEEE Transactions on Electron Devices **51**, 172 (2004).

[5] J. Guo, S. Datta, M. Lundstrom, and M. P. Anantram, International Journal of Multiscale Computational Engineering **2**, no. 2 (2004).

[6] A. Javey *et al.*, Nano Letters **4**, 1319 (2004).

[7] J. Guo and M. Lundstrom, Appl. Phys. Lett. **86**, 193103 (2005).

[8] L. D. Landau and E. M. Lifshitz, *Quantum mechanics: Non-relativistic Theory*, **3**rd edition (Butterworth-Heinemann, London, 1981), Chap. VII.

[9] J. Appenzeller, Y.-M. Lin, J. Knoch, and Ph. Avouris, Phys. Rev. Lett. **93**, 196805 (2004).

[10] S. Datta, *Quantum Transport: Atom to Transistor*, **2**nd edition (Cambridge University Press, Cambridge, 2005).

[11] R. Venugopal, Z. Ren, S. Datta, M. S. Lundstrom, and D. Jovanovic, J. Appl. Phys. **92**, 3730 (2002).

[12] S. Datta, *Electronic Transport in Mesoscopic Systems* (Cambridge University Press, Cambridge, 1995), Chap. 8.

[13] A. Svizhenko, M. P. Anantram, T. R. Govindan, B. Biegel, and R. Venugopal, J. Appl. Phys. **91**, 2343 (2002).

[14] N. S. Wingreen, K. W. Jacobsen, and J. W. Wilkins, Phys. Rev. Lett. **61**, 1396 (1988).

[15] L. I. Glazman and R. I. Shekhter, Solid State Commun. **66**, 65 (1988).